\documentclass[aps,a4paper,10pt,twocolumn,nofootinbib,floatfix]{revtex4} 
\usepackage{amsfonts}
\usepackage{amssymb}
\usepackage{mathrsfs}
\usepackage[mathscr]{euscript}
\usepackage[dvips]{graphicx}
\usepackage{fancyhdr}
\usepackage[hypertex=true]{hyperref} 
\usepackage{makeidx}
\usepackage{colordvi}

\begin{document}

\title{Proposal for absolute CEP measurement using $0$-to-$f$ self-referencing}

\author{S. B. P. Radnor}

\author{P. Kinsler}
\email{Dr.Paul.Kinsler@physics.org}

\author{G. H. C. New}

\affiliation{
  Blackett Laboratory, Imperial College London,
  Prince Consort Road,
  London SW7 2AZ,
  United Kingdom.}

\begin{abstract}

We show how to adapt
 a $0-f$ self-referencing technique
 \cite{Fuji-RGAUYTHK-2005njp,Rauschenberger-FHYUGHK-2006lpl}
 to provide a single shot absolute Carrier Envelope Phase (CEP) measurement
 by using the CEP reference provided by 
 difference frequency generation (DFG) between the spectral wings
 of the fundamental pulse.
Usually, 
 the beat between the input pulse and the DFG signal then provides
 feedback with which to stabilize the CEP slip in a pulse train.
However, 
 with a simple extension we can get a single shot absolute CEP measurement.
Success relies on having well characterized input pulses, 
 and the use of accurate propagation models through the nonlinear crystal --
 these enable us to construct a mapping between the experimental measurement
 and the CEP of the optical pulse.

\end{abstract}





\lhead{\includegraphics[height=5mm,angle=0]{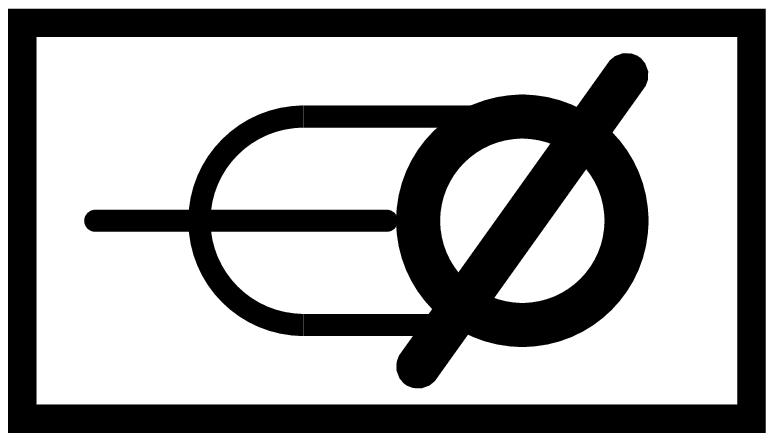}~~ABSCEP}
\chead{~}
\rhead{
\href{mailto:Dr.Paul.Kinsler@physics.org}{Dr.Paul.Kinsler@physics.org}}

\date{\today}
\maketitle
\thispagestyle{fancy}


\noindent
Note: 
 The research in this paper comprised
 part of S. B. P. Radnor's PhD dissertation \cite{Radnor-2007phd}, 
 and was completed in conjunction with the other authors.
It was initially reported in 2007 at ECLEO in Munich.
This text was eventually finalised by P. Kinsler.

\section{Introduction}
\label{S-Introduction}

The management of Carrier Envelope Phase (CEP)
 is of paramount importance in attosecond physics
 and frequency metrology.
In the absence of intervention, 
 the CEP shifts from one pulse to the next in a pulse train,
 a feature that is unacceptable in experminents involving few-cycle pulses.
The first step is to stablize the CEP, 
 the second to measure the CEP offset once stabilized, 
 and the third to be able to create any desired CEP to order.

Recently, 
 a $0-f$ self-referencing technique for CEP stabilization was developed
 \cite{Rauschenberger-FHYUGHK-2006lpl,Fuji-RGAUYTHK-2005njp}
 as an alternative to the $f-2f$ schemes \cite{Fortier-JC-2003ol}
 previously used for CEP stabilization.
The scheme was based on difference frequency generation (DFG) 
 \cite{Boyd-NLO}
 between the spectral wings of the input pulse, 
 a process in which the overall CEP is cancelled out.
The beat between the DFG signal
 and the extreme low-frequency wing of the pulse
 then provides feedback to the source laser to enable elimination 
 of the CEP drift.

Measurements of the absolute CEP 
 have been based on a number of techniques 
 including photo-ionization \cite{Apolonski-DPKHULTBHK-2004prl}
 high harmonic generation \cite{Haworth-CRKMT-2007np}, 
 and plasma generation \cite{Kress-LTDGZEMMUR-2006np}.
Some innovative methods for single-shot CEP measurement 
 based on spectral interference
 have also been suggested \cite{Mehendale-MLVC-2000ol,Kakehata-TKTFHT-2001ol}.
These methods rely on CEP dependent interference
 occurring between various harmonics. 
In the case of Mehendale et al. \cite{Mehendale-MLVC-2000ol}
 this involves interference between the second and third harmonics, 
 whereas the work by Kakehata et al. \cite{Kakehata-TKTFHT-2001ol}
 relies on interference between a delayed fundamental and its second harmonic.
Both mechanisms work on the basis of a relative CEP dependent relationship 
 being enforced by the nonlinearity. 
Though the schemes are interesting, 
 it is not clear how sensitive their interference assumptions
 are to intensity fluctuations, 
 propagation distance,
 and so on.

In the present paper, 
 we show how the $0-f$ technique 
 \cite{Rauschenberger-FHYUGHK-2006lpl,Fuji-RGAUYTHK-2005njp}
 can be extended to enable a measurement of absolute CEP to be made 
 \cite{Radnor-2007phd}.
Provided the evolution of the pulse
 within the nonlinear crystal used for the DFG
 can be accurately mapped, 
 we show that the CEP can be recovered from the detailed characteristics 
 of the beat signal used as the feedback source
 in the original experiment.
We test the robustness of this technique to phase and intensity variations.

In section \ref{S-CEP},
  we establish a rigorous definition of CEP
 that provides a sound basis for section \ref{S-Scheme}, 
 where we discover how the absolute CEP
 can be recovered from an interferometric measurement.
In section \ref{S-modelling},
 we describe the numerical techniques
 needed to extract the CEP value from the interference record, 
 and in section \ref{S-Testing} we test their reliability.
In section \ref{S-Measuring},
 we show how this enables one to measure the absolute CEP
 of an input pulse, 
 followed in section \ref{S-Conclude} by our conclusions.

\section{Carrier envelope phase}
\label{S-CEP}

Developments in ultrafast optical pulses have led to the production 
 of sub-cycle pulses.
In these limits, 
 robust definitions are needed to fully characterise the pulse, 
 as common descriptors can become ambiguous or fail. 
Perhaps the best example of this ambiguity
 is the representation of a pulse with a carrier and envelope, 
 where even as early as 1946 it was known
 that carrier envelope decompositions (of radar pulses) were not unique
 \cite{Gabor-1946jiee}.
Brabec and Krausz went some way towards dealing with these issues
 by suggesting a definition for the central frequency, 
 and stating that an envelope definition is only valid
 if it remains invariant under a change of phase \cite{Brabec-K-1997prl}.

The most natural way of defining CEP 
 would appear to be based on a time-domain picture
 in which the time interval between the peak of the pulse
 and the closest maximum or minimum
 is measured as a fraction of the optical period.
However, 
 for more complex pulse shapes
 there is no unambiguous way to determine peak of the pulse\footnote{It is
  of course possible to invent schemes for generating a suitable 
  centre position, 
  e.g. by calculating a weighted average
  over its intensity profile 
  (see e.g. \cite{Brabec-K-1997prl}) -- 
  however this can perform poorly for pulses with satellite peaks},
 or to determine the period
 when the spectral bandwidth is broad.
Further, 
 a fixed pulse envelope in concert with a varying carrier phase
 generates a pulse that does not guarantee a constant energy 
 or satisfaction of the the zero-force condition
 (see e.g. \cite{Milosevic-PBB-2006jpb}).
We therefore adopt a purely spectral approach 
 in which we define the absolute spectral CEP of a pulse
 using the equation
~
\begin{eqnarray}
  \Psi_0(\omega)
&=&
  \phi_0 
 +
  \psi(\omega)
.
\label{eqn-phase-single-pulse}
\end{eqnarray}
The CEP we wish to measure is $\phi_0$, 
 while the relative spectral phase $\psi(\omega)$
 is set to zero at a chosen centre frequency $\omega_0$.
Since $\psi(\omega_0)=0$, 
 it follows that $\phi_0=\Psi(\omega_0)$.
We note that $\psi(\omega)$ 
 is a measurable quantity even for ultrashort pulses -- 
It is possible to determine the relative spectral phases
 of few-cycle pulses 
 \cite{Cheng-FSLSK-1999ol,Kobayashi-SF-2001ieeqe}\footnote{In 
   Kobayashi et al.'s Fig. 4(b), 
   $\psi$ extends from 550nm to 800nm.
   In our scheme,
   $\psi$ will most likely need to be characterised 
   further down into the low frequency wing -- 
   our $\omega_{+}/\omega_d$ ratio is $2.6$; 
   for Kobayashi et al. the ratio was only $1.5$.}.
This can be achieved to an accuracy of 0.04 rads 
 using the SPIDER technique \cite{Anderson-AKW-2000apb}.
This $\psi(\omega)$ tells us the relative CEP
 of all frequency components of the pulse, 
 something which we use in the scheme presented in this paper.

Having considered a single pulse, 
 we now need to consider a pulse train, 
 the spectrum of which is a comb
 of equally spaced frequency components
 with the lowest frequency tooth at $\omega_{cep}$.
The phase slip between pulses is $\Delta = \omega_{cep} \tau$,
 where $\tau$ is the time interval
 between sucessive pulses in the train.
Hence the \emph{absolute} spectral phase of the $n$-th pulse in a train is
~
\begin{eqnarray}
  \Psi_n(\omega)
&=&
  \phi_n
 +
  \psi(\omega)
\label{eqn-phase-nth-pulse}
,
\end{eqnarray}
where $\phi_n = \phi_0 + n \Delta$.

%
\section{Scheme}
\label{S-Scheme}

If we know the relative spectral phase $\psi(\omega)$, 
 \emph{and} we can determine the absolute CEP of any one frequency,
 then we can calculate the absolute CEP of any other.
The advantage of a DFG process \cite{Boyd-NLO}
 is that it provides us with just such a reference CEP.

\begin{figure}[ht]
\centering
\includegraphics[width=0.70\columnwidth]{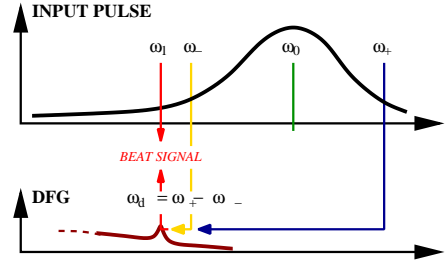}
\caption{Diagram of the DFG signal being generated, 
 comparing the phase matched DFG component to 
 the wing of the input pulse.
}
\label{fig-diagram}
\end{figure}

To get efficient DFG, 
 a nonlinear crystal (such as MgO:LN)
 is periodically poled to phase match
 a particular frequency mixing process.
The poling period is chosen 
 so that selected frequencies on the 
 upper ($\omega_{+}$) and lower ($\omega_{-}$) wings
 of the pulse spectrum
 are phase matched for DFG
 at $\omega_d=\omega_{+}-\omega_{-}$, 
 as illustrated on fig. \ref{fig-diagram}.
The nonlinear interaction generates a polarization term
 whose phase $\Psi_{p}$ is 
~
\begin{eqnarray}
  \Psi_{p}
&=&
  \Psi_n(\omega_{+})
 -
  \Psi_n(\omega_{-})
~~~~
= 
  \psi(\omega_{+})
 -
  \psi(\omega_{-})
.
\label{eqn-Psi-polariz}
\end{eqnarray}
Since the unknown offsets $\phi_0$ and $\Delta$ have canceled out, 
 $\Psi_{p}$ can be calculated from $\psi(\omega)$, 
 and provides a CEP reference.
Unfortunately, 
 this cannot be measured directly, 
 so we have to analyse how 
 both the incident pulse 
 and the DFG propagates through the crystal,
 and how they interfere and are measured.

\emph{First,} 
 let us consider the DFG component.
The polarization term (with absolute phase $\Psi_{p}$) will then 
 generate a DFG signal that \emph{exits} the crystal
 with a phase shifted by an amount $\delta_p$, 
 so that 
~
\begin{eqnarray}
  \Psi_{d}
&=&
  \Psi_{p} + \delta_p
.
\label{eqn-Psi-DFG}
\end{eqnarray}

We can see how this shift $\delta_p$ arises
 in the idealised case where a detectable DFG signal
 could be generated from a thin layer 
 of dispersionless $\chi^{(2)}$ medium.
Here $\delta_p=\pi/2$, 
 since the DFG signal is just the integral of the 
 driving polarization.
In general, 
 however,
 $\delta_p$ will be a complicated function 
 of both $\Psi_{p}$ and the pulse intensity and profile, 
 since it results from propagation through a crystal which is 
 both nonlinear and dispersive.

\emph{Second,} 
 we need to consider the wing of the 
 input pulse spectrum at frequency $\omega_d$, 
 which is co-propagating with the DFG component discussed above.
This part of the pulse had an initial phase $\Psi_n(\omega_{d})$, 
 but this changes as it propagates through the crystal, 
 and on exit it has a phase $\Psi'_n(\omega_{d})$
 that has been shifted by $\delta_d$,
 i.e. 
~
\begin{eqnarray}
  \Psi'_n(\omega_{d}) 
&=&
  \Psi_n(\omega_{d}) + \delta_d
.
\label{eqn-Psi-propagate}
\end{eqnarray}
Although $\delta_d$ will predominantly arise from the dispersion,
 but there will also be additional contributions,
 due to other DFG processes 
 (i.e. those not involving $\omega_{+}, \omega_{-}$), 
 from SPM, four-wave mixing or other processes.
However, 
 with careful design these can be minimized, 
 and so will only add an unimportant 
 (but nevertheless calculable) offset to $\delta_d$.

\emph{Lastly, }
 we apply our understanding of both the DFG and pulse propagation
 to determine the interference between them, 
 and what would be measured on a photodetector.
As the pulse exits the crystal, 
 the DFG component (with phase $\Psi_{d}$) 
 will interfere with the wing of the input pulse 
 at that same frequency, 
 which now has phase $\Psi'_n(\omega_{d})$.
Thus the photodetector sees an interference
 between the DFG and the spectral wing at frequency $\omega_d$.
From this interference it is possible to 
 infer the relative phase $\delta_i$ between the contributions, 
 which is
~
\begin{eqnarray}
  \delta_i
&=&
  \Psi'_n(\omega_{d})
 - 
  \Psi_{d} 
\\
&=&
  \Psi_n(\omega_{d})
 +
  \delta_d
 - 
  \Psi_{d} 
.
\label{eqn-Psi-interference}
\end{eqnarray}
In order to optimize the visibility of this interference, 
 we need to ensure that the DFG signal and the amplitude of the 
 wing of the input pulse at that frequency are comparable; 
 if either is too dominant, 
 the CEP sensitive modulation of the interference will be less detectable.

In existing $0-f$ CEP stabilization 
 experiments\cite{Rauschenberger-FHYUGHK-2006lpl,Fuji-RGAUYTHK-2005njp}, 
 the evolving interference signal 
 resulting from the train of CEP-slipping pulses
 produces a beat signal dependent on the CEP slip.
This beat is then used in feedback designed to reduce
 the CEP slip to zero, 
 thus \emph{stabilizing} the CEP of the pulses in the train
 to a fixed (but unknown) value $\Psi_n(\omega)=
  \phi_0  +  \psi(\omega)$.

\emph{The scheme presented here works 
 because we incorporate additional information 
 based on knowledge of how 
 the pulse propagates through the crystal.}
This means that 
 we can predict 
 the phase shifts $\delta_p$ and $\delta_d$, 
 at which point the interference measurement (i.e. of $\delta_i$)
 turns a  simple CEP stabilization
 into a CEP measurement.

To make it clear how the CEP measurement is constructed, 
 we now show how the various phases in the scheme are related.
We are specifically interested in the CEP $\phi_0$ 
 of the central frequency $\omega_0$ of the pulse; 
 so it is useful to write down how 
 the relative phases between the low frequency ($\omega_d$) wing
 of the input pulse and 
 its centre are
~
\begin{eqnarray}
  \Psi_n(\omega_0)
&=&
  \Psi_n(\omega_{d})
 -
  \psi(\omega_{d})
.
\end{eqnarray}
Now, 
 by substituting in the preceeding collection of phase relationships
 (eqns. (\ref{eqn-Psi-polariz}, \ref{eqn-Psi-DFG},
         \ref{eqn-Psi-propagate}, \ref{eqn-Psi-interference})), 
 we can get
~
\begin{eqnarray}
  \Psi_n(\omega_0)
=
  \phi_n
&=&
  \psi(\omega_{+})
 -
  \psi(\omega_{-})
 +
  \delta
 -
  \psi(\omega_{d})
,
\label{eq-absphi}
\end{eqnarray}
where the sum of all the nonlinearity and propagation 
 phase shifts is $\delta = \delta_{p} + \delta_{i} - \delta_{d}$.

Thus if we 
 [a] know the relative phase spectrum $\psi(\omega)$ of the input pulse(s), 
 [b] understand the phase shifts introduced by propagation 
 through the crystal ($\delta_d$) and of the DFG ($\delta_p$),
 and then can 
 [c] \emph{measure} the interference phase difference ($\delta_{i}$), 
 we will know every part of the RHS of eqn. (\ref{eq-absphi}) --
 i.e. we know the absolute CEP $\phi_n$ of the incoming pulse.

The most challenging part of the CEP measurement is 
 determining the $\delta_{p}$ and $\delta_{d}$ contributions.
Fortunately, 
 we can avoid having to calculate them individually 
 by calculating them all at the same time in a numerical simulation, 
 as discussed in the following section.

%
\section{Modeling}
\label{S-modelling}

The modeling is a crucial part of the scheme, 
 since it allows us to determine how differing input CEPs
 map onto the detected interference measurements.
After choosing our crystal and evaluating its parameters, 
 and characterising the pulses in our pulse train 
 (particularly $\psi(\omega)$, 
 we run a set of simulations over the range of CEPs.
The results can then be used 
 to build a map between the input CEP and the interference signal.
To do this we need to take the spectrum of each output pulse
 from a simulation, 
 and integrate over the detector response.
In our results, 
 we assume the spectral response of an InGaAs-Hamamatsu PD 
 as used by Fuji and others 
 \cite{Fuji-RGAUYTHK-2005njp,Rauschenberger-FHYUGHK-2006lpl}.
We then need to check the mapping, 
 and ensure that we will get the required level of discrimination 
 between interference measurements from pulses with different CEP's.

To do the propagation part of the modeling, 
 we solve Maxwell's equations
 using the PSSD technique \cite{Tyrrell-KN-2005jmo}
 for a chosen crystal thickness. 
In existing $0-f$ experiments, 
 the crystal thickness is typically of the order of millimetres, 
 so diffraction is negligible. 
We consider MgO:LN, 
 with parameters taken from 
Further, 
 with a suitable choice of nonlinear crystal 
 (i.e. MgO:LN, 
 and parameters from \cite{Handbook-NLOCX,Zheng-WPCF-2002josab}),
 the nonlinear $\chi^{(2)}$ interaction occurs only in the 
 extraordinary polarization (i.e. is $e+e\rightarrow e$), 
 allowing us to further reduce Maxwell's equations to
~
\begin{eqnarray}
\label{eq:Max}
  \frac{\partial E_{x}}{\partial z}
&=&
 -\mu_{0}\frac{\partial H_{y}}{\partial t}
\\
  \frac{\partial H_{y}}{\partial z}
&=&
 -
  \epsilon_{0}\frac{\partial}{\partial t}\left[E+\chi^{(1)}*E
 +
  \chi^{(2)}E^{2}
 +
  \chi^{(3)}E^{3}\right]
,
\label{eqn-maxwellseqns}
\end{eqnarray}
where $\chi^{(1)}$ contains linear dispersion, 
 and any nonlinear response is assumed to be instantaneous.

In existing experiments
 \cite{Rauschenberger-FHYUGHK-2006lpl,Fuji-RGAUYTHK-2005njp}, 
 a MgO:LN crystal is periodically poled at $11.21\mu$m, 
 and is optimized for DFG between the wings of the fundamental: 
 $\omega_{+}(3.04\times 10^{15})-\omega_{-}(1.885\times 10^{15})
 =\omega_{d}(1.155\times 10^{15})$ [rad s$^{-1}$]. 
The peak pulse power was calculated
 to be $\simeq5\times 10^{11} \textrm{W/cm}^{2}$, 
 with a duration of $\sim$ 6fs (830 nm carrier).
At these pulse powers in MgO:LN, 
 the relative nonlinear strengths in the crystal at the 
 pulse peaks are 
 $\chi^{(2)} E = 0.082$
 and $\chi^{(3)} E^{2} = 0.0027$. 
This means that the self-phase modulation (SPM) distance is $L_{SPM}=0.26$mm, 
 implying significant SPM over a 2mm crystal, 
 along with other $\chi^{(3)}$ effects such as 4-wave mixing, 
 further complicating propagation and DFG.
Modelling of extreme SPM (only) and sensitivity to CEP 
 has been considered by Kinsler \cite{Kinsler-2007-phspm}
 and also Genty et al.\cite{Genty-KKD-2008oe}.

With these parameter values, 
 the pulses inside the crystal decohere rapidly
 because of their high intensity and short duration.
In combination 
 with the relatively small energy content within the spectral wings, 
 the DFG does not continue to grow throughout the whole crystal, 
 but does so only over several coherence lengths. 
Nevertheless, 
 the DFG signal can still be made large enough for the 
 photodiode (PD) to detect a beat against the wing of the input spectrum.

In our modeling, 
 we retained the bulk of these parameter values to match experiment, 
 but adjusted the crystal thicknesses to $\sim 100\mu$m.
This is because thinner crystals still create sufficient DFG, 
 but also have the advantage
 of producing a cleaner CEP to interference signal mapping.

%
\section{Testing the CEP response at $\omega_{d}$}
\label{S-Testing}

In order to make our scheme work,  
 we need to guarantee that the response of the measured
 interference signal depends on the CEP 
 in a reliable way, 
 \emph{and} that it is sufficiently insensitive to other 
 pulse characteristics, 
 such as intensity.
Obviously these will depend on the particular design 
 of the experiment, 
 e.g. the size of the chosen nonlinear crystal, 
 the periodic-poling length, 
 pulse frequency, 
 and so on.
In this section, 
 we use the parameters described in the previous section 
 to test the stability of the pulse propagation and interference signal 
 against CEP variation and intensity fluctuations.

%
\subsection{Response to CEP slip $\Delta$}
\label{Ss-Testing-ce}

\begin{figure}[ht]
\centering
\includegraphics[angle=-90,width=0.70\columnwidth]{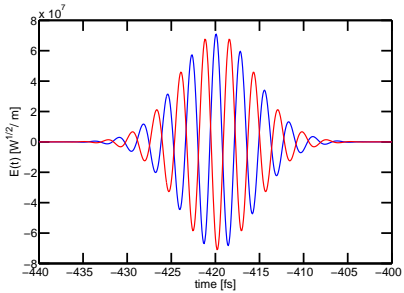}
\caption{
Initial pulses with $\Psi_n(\omega_0)=0$ and $\pi/2$ (cosine \& sine).
}
\label{fig-CEP_10}
\end{figure}

To test whether the interference signal 
 will behave as expected, 
 we did a set of simulations.
Each simulations started with the same parameters, 
 except for a cumulative
 ``shot-to-shot'' CEP slip of $\Delta = \pi/10$.
Fig.  \ref{fig-CEP_10} shows example input pulses.

Let us consider an idealized case, 
 with only DFG and a linearly propagated input pulse present.
Here we should see two regimes when comparing the 
 phase spectrum of output pulses
 generated by input pulses differing by a CEP slip $\Delta$.

\begin{enumerate}

\item
The spectrum of the input pulse dominates the contribution from the DFG signal, 
 typically this occurs closer to the centre of the pulse spectrum, 
 i.e.  $\omega>\omega_{1}$
In this case, 
 the CEP of the pulse dominates, 
 so that the phase difference between subsequent pulses 
 will just be the inter-pulse CEP slip $\Delta$.

\item
The DFG signal dominates the contribution from the input pulse, 
 typically this occurs for low frequencies, 
 i.e. $\omega<\omega_{1}$.
Since the DFG is insensitive to the input CEP, 
 the phase difference between subsequent pulses 
 will be zero.

\end{enumerate}

\begin{figure}[ht]
\centering
\includegraphics[angle=-0,width=0.70\columnwidth]{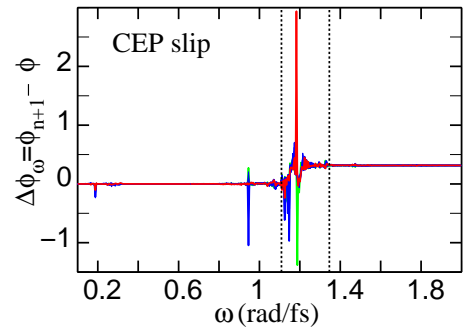}
\caption{Plot of the shot-to-shot changes of phase. 
The right hand side of the figure $\omega>\omega_{1}$ shows
 the expected $\pi/10$ phase change, 
 while the $\omega<\omega_{1}$ region (left)
 represents the phase stable signal. 
The noise present at $\omega\simeq\omega_{1}$ is the interference
 between $\omega_{1}$ and $\omega_{pm}$. 
The dotted lines represent the spectral range of the PD 
 (colours: green=initial, red=final and blue
 are the intermediate $\Delta\phi_{\omega}$ curves). 
}
\label{fig:cep}
\end{figure}

In between these two regimes
 will be a transition region where the two are comparable; 
 this is just the regime in which we look for the interference
 between the incoming pulse 
 (with its phase slip $\Delta$, as described in point 1 above);
 and the phase stabilised DFG 
 (with no phase slip, as described in point 2 above).
Both regions, 
 and the transition region of interference between them 
 can be seen on fig. \ref{fig:cep}.
It is important to note that, 
 the change between different pairs of simulations
 is small -- 
 irrespective of the absolute CEP values chosen for the two pulses.
The only departures are at the narrow spikes
 caused by the (expected) strong CEP sensitivity
 near the nodes of the pulse.
As an aside, 
 we could (if desired) also estimate the linearity
 of the response to the CEP slip
 as done by Kinsler \cite{Kinsler-2007-phspm} 
 in the extreme nonlinear regime; 
 that work also suggests more systematic tests of simulation pairs
 to investigate the CEP dependence.

Here, 
 however, 
 we are satisfied by the fact that fig. \ref{fig:cep} not only shows
 the predicted DFG phase stabilized region at low frequencies, 
 but also the existance of a transition region where strongly modulated 
 phase sensitive intereference takes place.

%
\subsection{Response to intensity fluctuations}
\label{Ss-Testing-I}

Our scheme relies on nonlinear interactions, 
 but these are strongly intensity dependent.
This means that intensity changes between pulses in the train
 might change the interference signal
 in a way that masks the CEP sensitivity
 of the interference signal.
We now test the effect of intensity fluctuations
 by fixing the CEP for a set of simulations, 
 whilst making shot-to-shot changes in intensity
 spanning a range of $\sim \pm 1$\%.

The results of this simulation set are
 displayed on fig. \ref{fig:cep}, 
 which demonstrates that intensity variation
 has a relatively weak effect in the interference region.
This means that the mapping between photodetector signal
 and absolute CEP will be \emph{insensitive}
 to the intensity fluctuations in the pulse train.
As a result, 
 for our parameters, 
 we can disregard the effect of intensity fluctuations
 when it comes to reconstructing the CEP of an individual pulse.
In a more extreme nonlinear case this is not always true, 
 see e.g. \cite{Kinsler-2007-phspm}; 
 but here the nonlinear phase shifts 
 (e.g. those due to SPM)
 are not strongly intensity dependent.

\begin{figure}[ht]
\centering
\includegraphics[angle=-0,width=0.70\columnwidth]{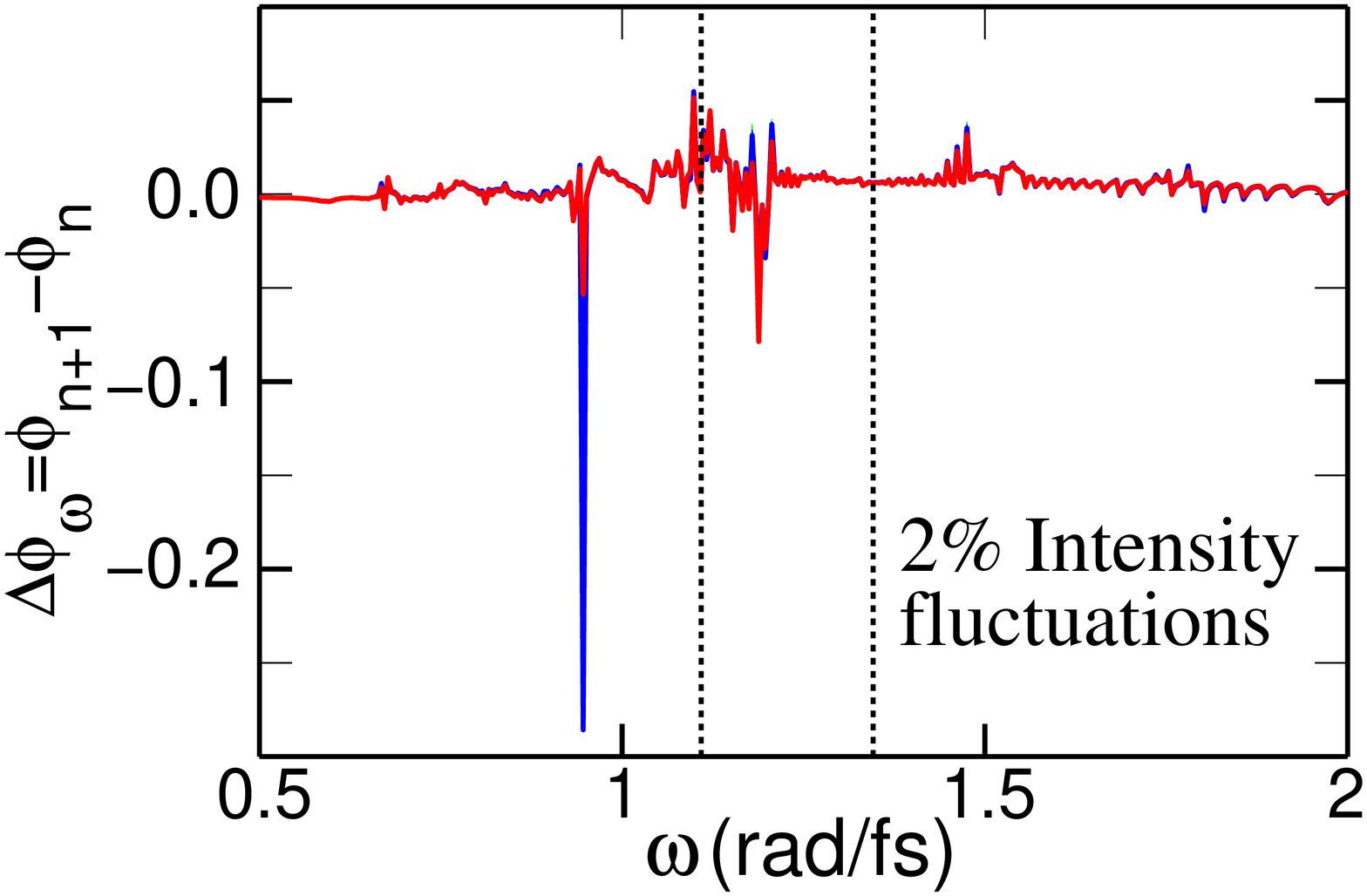}
\caption{Plot of $\Delta\phi_{\omega}$ for intensities
 ranging from 99$\%$ to 101$\%$. 
All pulses began with $\phi_{CE}=0$. 
The red and green curves represents the lowest and highest intensities respectively. 
The 2$\%$ range is divided into 10 equal sections (blue). 
}
\label{fig-intensity}
\end{figure}

%
\section{Measuring the Absolute CEP}
\label{S-Measuring}

In the previous section we demonstrated that 
 (or our chosen parameters) 
 not only was the interference in the $\omega_{d}$ DFG region 
 sensitive to CEP in a controllable way, 
 the effect would not be masked by intensity fluctuations.
Since we see this clear dependence on CEP, 
 it is possible to determine the absolute CEP from the interference signal --
 as long as the pulse intensity and crystal parameters
 are appropriately matched.
For example, 
 the CEP dependence is better behaved at some distances than at others -- 
 for our chosen parameters, 
 it happens that distances of $\sim 50, 100\mu$m give good results.

\begin{figure}[ht]
\centering
\includegraphics[angle=-90,width=0.70\columnwidth]{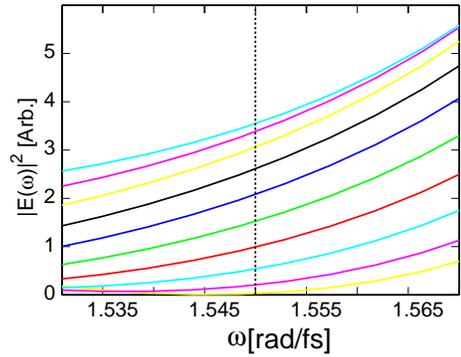}
\caption{
Plot of the spectral intensity at $\omega\simeq\omega_{pm}$ 
 for a propagation distance of $50\mu$m.
The CEP dependent structure can clearly be seen,
 where the lowest curve (yellow) corresponds to $\phi_{CE}=0$
 and the upper (turquoise) curve corresponds to $\phi_{CE}=9\pi/10$. 
The range is divided into intervals of $\pi/10$ rads. 
}\label{fig:wow1}
\end{figure}

\begin{figure}[ht]
\centering
\includegraphics[angle=-90,width=0.70\columnwidth]{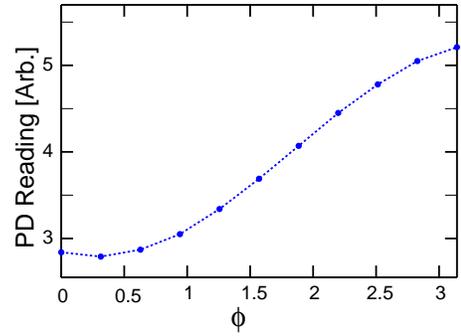}
\caption{
Mapping between photodetector signal and pulse CEP: 
By integrating the power falling on the PD within the detectable range, 
 a CEP varying signal was constructed. 
In this case $\phi=0$ maps to the first point on the PD signal
 (an extra point has been plotted for $\phi_{CE}=\pi$).
}\label{fig:wow2}
\end{figure}

Fig. \ref{fig:wow1} shows the CEP dependent structure at 
 a propagation distance of 50$\mu$m, 
 where each curve is the intensity for a different input CEP.
Using this, 
 we can then integrate that spectral behaviour
 over the response of the photodetector to generate 
 our mapping.
The mapping corresponding to fig. \ref{fig:wow1}
 and our chosen photodiode is shown on fig. \ref{fig:wow2}.

However, 
 because each interference signal value is not unique, 
 we have only determined the CEP to within $\pi$. 
To complete the determination of the CEP to within a $2\pi$ range
 we need to take two such measurements under slightly different conditions, 
 e.g. using different propagation lengths.

Summarizing,
 for this CEP measurement scheme to succeed,
 we must have accurate knowledge 
 of the following three things:

\begin{enumerate}

\item \textbf{ The phase spectrum $\psi(\omega)$ of the pulse.} 
In eqn.(\ref{eq-absphi}) we see that it depends on the sums or difference
 of four values of $\psi(\omega)$, 
 thus compounding any uncertainty in its determination.

\item \textbf{ The intensity fluctuations of the pulse. }
The intensity must be controlled sufficiently well so that the SPM 
 or XPM effects are minimized,
 and do not significantly alter the propagation of the pulse through
 the crystal.

\item \textbf{ Pulse propagation.} 
This can be done easily with PSSD simulation code (or similar), 
 but we need accurate information on the initial conditions
 of the pulses in the pulse train.

\end{enumerate}

Remarkably, 
 this can be done with a small extension to current $0-f$
 self-referencing methods, 
 which are currently only used to stabilize the CEP.
The extension is to numerically model the propagation of the 
 pulse through the nonlinear crystal in order to determine 
 the mapping between the photdetector signal and the input CEP.

%
\section{Conclusion}
\label{S-Conclude}

We have demonstrated how an ordinary $0-f$ self-referencing scheme
 can be easily extended to measure absolute CEP,
 rather than just being used as a CEP stabilization tool. 
The scheme relies on a phase stable signal being passively 
 produced through DFG, 
 and does not require strong field physics to operate.
Instead, 
 we propose to numerically model the propagation of the 
 pulse through the nonlinear crystal; 
 and to use the information gained 
 to determine the mapping between the detected interference signal
 and absolute CEP of the input pulse.

%

\bibliography{/home/physics/_work/bibtex.bib}

\end{document}